\documentclass[conference]{IEEEtran}
\IEEEoverridecommandlockouts
\usepackage{cite}
\usepackage{amsmath,amssymb,amsfonts}
\usepackage{algorithmic}
\usepackage{graphicx}
\usepackage{textcomp}
\usepackage{multirow}
\usepackage{amssymb}
\usepackage{booktabs}
\usepackage{threeparttable}
\usepackage{enumitem}
\usepackage[framemethod=tikz]{mdframed}
\usepackage{tikz}
\usepackage{tipa}
\usepackage{xtab}
\usepackage{mfirstuc}
\usepackage{todonotes}
\usepackage{soul}
\usepackage{colortbl}
\usepackage{xspace}
\usepackage{xcolor}
\newcommand{\hghlt}[1]{\textcolor{blue}{#1}}
\usepackage{url}

\usepackage{caption}
\usepackage{subcaption}
\usepackage[many]{tcolorbox}
\newtcolorbox{mybox}[1][]{
  breakable,
  title=#1,
  colback=white,
  colbacktitle=white,
  coltitle=black,
  fonttitle=\bfseries,
  bottomrule=0pt,
  toprule=0pt,
  leftrule=3pt,
  rightrule=3pt,
  titlerule=0pt,
  arc=0pt,
  outer arc=0pt,
  colframe=black,
}

\newcommand{\catname}[1]{\textsf{\small #1}}

\newcommand{\qs}[2]{\emph{``#1'' (#2)}}
\newcommand{\qsd}[3]{\emph{``#1'' (#2 on #3)}}

\def\BibTeX{{\rm B\kern-.05em{\sc i\kern-.025em b}\kern-.08em
    T\kern-.1667em\lower.7ex\hbox{E}\kern-.125emX}}
\begin{document}

\title{Narratives: the Unforeseen Influencer of Privacy Concerns\\

}

\author{\IEEEauthorblockN{
Ze Shi Li, 
Manish Sihag,
Nowshin Nawar Arony,
Joao Bezerra Junior,
Thanh Phan,
Neil Ernst, 
Daniela Damian}
\IEEEauthorblockA{
\emph{Department of Computer Science}\\
\emph{University of Victoria, Victoria, Canada}\\
\emph{\{lize, manishsihag, nowshinarony, joaobatistajr, thanhpc, nernst, danielad\}@uvic.ca}}}

\maketitle

\thispagestyle{plain}
\pagestyle{plain}

\begin{abstract}
Privacy requirements are increasingly growing in importance as new privacy regulations are enacted.
To adequately manage privacy requirements, organizations not only need to comply with privacy regulations, but also consider user privacy concerns.
In this exploratory study, we used Reddit as a source to understand users' privacy concerns regarding software applications.
We collected 4.5 million posts from Reddit and classified 129075 privacy related posts, which is a non-negligible number of privacy discussions. 
Next, we clustered these posts and identified 9 main areas of privacy concerns.
We use the concept of narratives from economics (i.e., posts that can go viral) to explain the phenomenon of what and when users change in their discussion of privacy. 
We further found that privacy discussions change over time and privacy regulatory events have a short term impact on such discussions.
However, narratives have a notable impact on {what} and {when} users discussed about privacy.
Considering narratives could guide software organizations in eliciting the relevant privacy concerns before developing them as privacy requirements.

\end{abstract}


\section{Introduction}
Privacy is becoming, arguably more than ever, a critical non-functional requirement for software products and development organizations. 
Its importance is evident from the recent enactment of privacy regulations in jurisdictions worldwide \cite{radley-gardner_fundamental_2016, ccpa_california_2018, harding_five_2021, kadish_brazils_2020}. 
Non-compliance with these privacy regulations can result in heavy penalties for software organizations \cite{radley-gardner_fundamental_2016}. 
In addition, organizations are also amenable to users who are more vocal than ever with concerns related to privacy breaches and infringements that captured global attention \cite{ng_equifax_2017, silverstein_hundreds_2019, winder_235_2020}. 
User concerns about their privacy are warranted and understandable.
After all, serious personal data such as personally identifiable information or credit card information can be exposed when software fails to adequately protect or purposefully misuses personal data. 

Although seemingly separate, privacy regulations and user concerns are in fact interrelated. For example, recent privacy regulations such as the General Data Protection Regulation (GDPR) and California Consumer Privacy Act (CCPA) are impacting user privacy concerns. For an organization that must comply with privacy regulations, it is important that it recognizes and complies with not only regulatory mandated requirements, but also shared user concerns. 
This is however challenging for many organizations, particularly small resource constrained ones, that struggle with privacy compliance and must strike a delicate balance between regulatory compliance and business requirements. 
Consequently, user involvement and feedback is becoming an important avenue for organizations to identify areas of privacy concerns that users may have about their software, as well as privacy requirements that could be developed to address these concerns \cite{li_towards_2022}. Determining {when} and {what} concerns users express about privacy can be critical success factors for an organization's privacy measures.
Requirements engineering research has recognized the value of monitoring user feedback for the development of product requirements, though advances had been largely in identifying 
user feedback and functional requirements from app stores and forums \cite{tizard_can_2019, pagano_user_2013, williams_mining_2017}. 

Furthermore, user concerns may be influenced by \emph{Narratives} -- a concept known in economics \cite{shiller_narrative_2017, wydick_bruce_how_2015}, \cite{nerrative-defination} as  ``a song, joke, theory, explanation, or plan that has emotional resonance and that can easily be conveyed in casual conversation" \cite{shiller_narrative_2020}. 
With the increase of social media platforms engaging millions of users, stories are being shared and retold thousands of times forming different narratives \cite{page_introducing_2018}, which further impacts the user discussions on platforms.
Thus, narratives can be leveraged to identify the user feedback related to a software product.

\hghlt{} 
We report from an empirical study of Reddit as it allows for rich discussion between users in online communities. In Reddit each community is known as a ``subreddit" and allows users to engage in online discourse about a specific topic. 
A software developing organization can benefit from monitoring their associated subreddit and eliciting potential requirements from user feedback. 

Posts in a Reddit software forum have a higher character limit for their discussion as compared to app store reviews. 
This allows users to provide a more detailed review or longer discussion that covers multiple facets about a software, particularly non-functional requirements like privacy, which may be difficult to describe succinctly.

Hence, our study was guided by three research questions:
\begin{enumerate}[label=\textbf{RQ\arabic*},leftmargin=*]
    \item To what extent do users discuss privacy in software product forums?
    \item What privacy concerns do users discuss in software product forums such as Reddit?
    \item How do the number of privacy discussions change over time? 
\end{enumerate}

We collected 4.5 million posts from Reddit and used machine learning to classify whether a post is privacy or non-privacy related. We identified 129,075 privacy related posts and which, by applying clustering techniques, revealed 9 main areas of privacy concerns such as ``privacy issues and recommendations" and "privacy policy and permissions" among others. We then mapped these posts over time to see how privacy discussions changed over time, and sought reasons for variations in user concerns.

Our process of arriving at narratives was started as part of our preliminary analysis on the privacy posts and examination of outlier behaviour in user discussions over time.
We investigated if the privacy regulatory events could be the probable reason behind these trends.
However, these events could not explain the outlier behaviour in the user discussion. 
To further understand the factors that impacted such outlier behaviour in posts related to a software product we conducted a manual analysis and identified narratives, a concept that has been studied in fields like economics, to explain a potential impact on the outlier behaviour of posts. 
We conducted a top-down as well as a bottom-up approach to explore if narratives do have an impact on the discussions. 
Through this exploratory analysis (Telegram vs WhatsApp and Firefox vs Chrome), we hypothesized that narratives surrounding software products could play a role in shaping the user perception eventually impacting the user count on particular software products and draw implications for software developing organizations. 

Our work brings the following empirical contributions:
\begin{itemize}
    \item automated identification of privacy concerns in software product forums through machine learning; 
    \item empirical evidence on categories and temporal trends of privacy concerns in online communities in Reddit;
    \item hypothesizes the influence of narratives on privacy concerns,  which may differ from that of the more objective privacy policies.
\end{itemize}




\section{Methodology}
\label{methodology}
\begin{figure}[t]
    \centering
    \includegraphics[width=0.48\textwidth]{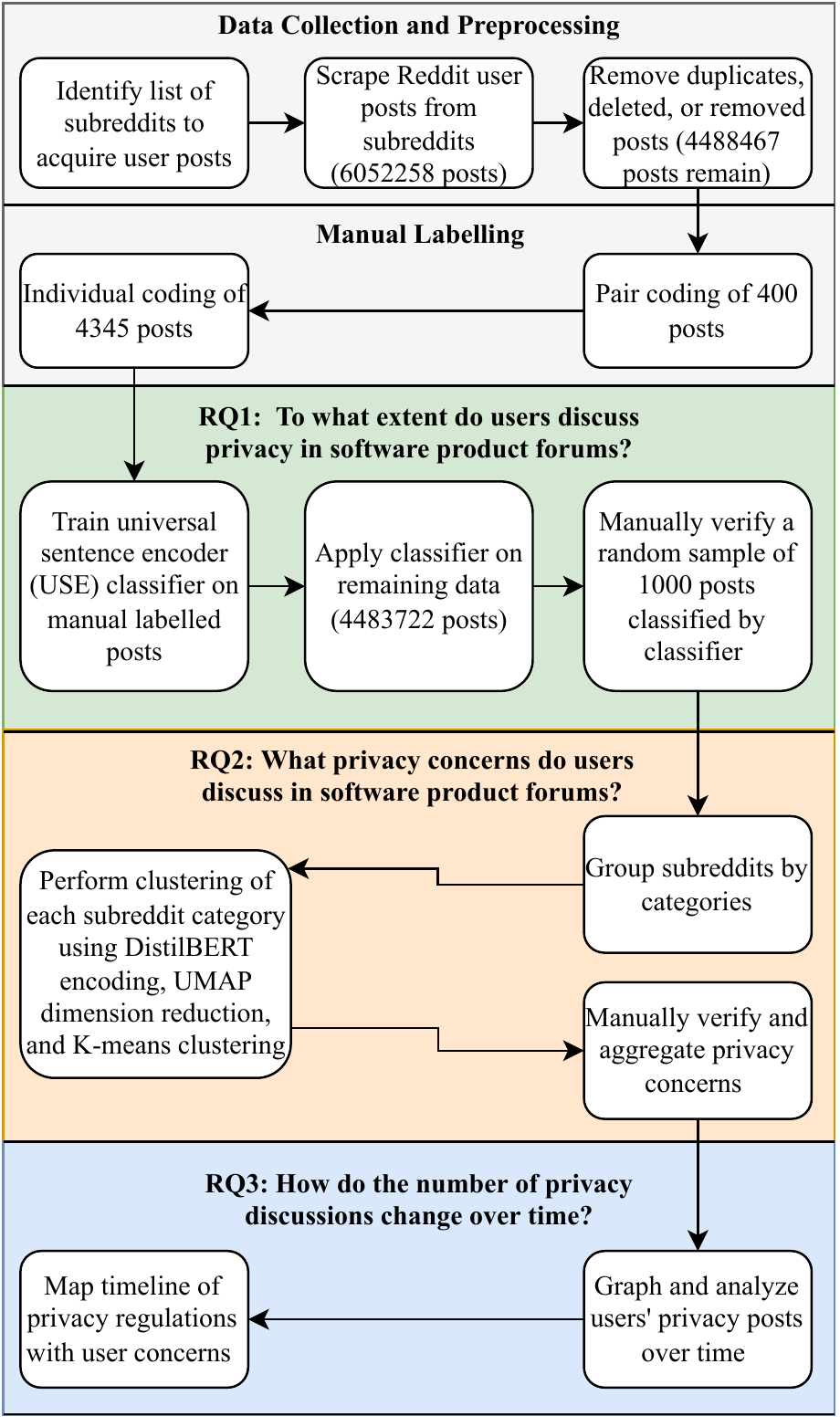}
    \caption{Research Process}
    \label{fig:research_methodology}
\end{figure}

We conducted an exploratory study on user privacy concerns and changes in privacy discussion over time in software product related user communities using Reddit as our source. 
Our methodology is summarized by Figure \ref{fig:research_methodology}.

\subsection{Data Collection}
To answer our RQs we first collected data from 66 online communities (i.e., subreddits). 
62 of the 66 are communities that are associated with popular software products such as WhatsApp, Telegram, AirBnb, or Instagram. 
We selected these software related subreddits because these subreddits should contain user discussions about the relevant software products. 
For example, if a user had a complaint or concern with a software like WhatsApp, they should in theory expound such issues in the WhatsApp subreddit. 
As our study was exploratory, we selected subreddits of large, well-known software products. 
We did not select subreddits for smaller, niche-specific software and we note that these smaller software may be a source for a future study. 
We identified a further four privacy specific subreddits, as we thought they might contain many user discussions about privacy regulations that we could analyze for RQ3.

We collected each subreddit's posts via Pushshift \cite{pushshift-github}, which is a large data store of all Reddit posts and comments and has a public facing API that supports ease of downloading for data analysis.
In total, we collected 6,052,258 posts across the 66 subreddits. 
For the 62 software product related subreddits, we collected 5,890,182 posts, and for the 4 privacy related subreddits, we collected 162,076 posts. We looked only at the original (possibly edited) post, not followup comments.
For a complete set of subreddits, please visit our replication package at \url{https://doi.org/10.5281/zenodo.6272629}.

After collection, we filtered the data to collect all posts between the creation of the subreddit and December 2021.
We removed any duplicate posts and any post that was either deleted by the author or removed by a moderator of the subreddit. 
Our data was left with 4,488,467 posts and we combined the title and text of each post to use for classification.

\subsection{Manual Labelling}
We manually labelled a ground truth dataset for training.
We collated a list of privacy terms that we believe would be relevant such as ``GDPR", ``CCPA", ``privacy" ,``leak", ``consent" and ``unlawful", \cite{radley-gardner_fundamental_2016} and ran each post against the list of privacy terms.
Based on the privacy term count, we randomly sampled 1721 0-privacy-term posts and 3024 1-or-more-privacy-term-posts.
Two of the co-authors of this paper, with experience in privacy requirements and requirement elicitation in industry, performed four rounds of pair coding before individually coding the rest of the posts.
The intent of pair coding was to establish a shared understanding of what it meant for a post to be privacy or non-privacy related. 
During pair coding, the inter-rater agreement were 77\%, 89\%, 83\%, 85\% and the Cohen's Kappa were 0.39, 0.53, 0.51, and 0.60, which our agreement levels hovered around moderate levels of agreement \cite{landis1977measurement}. 

Subsequent to the pair coding, the two co-authors individually coded a further 4345 posts.  
Of the manually labelled 4745 posts, 794 posts were labelled as privacy related and 3951 were labelled as non-privacy related. 
Since a user may write a post containing several critiques including, but not limited to, privacy, we labelled a post as \textit{privacy related} as long as part of the post referred to privacy. 
For example, \qs{I switched from my former browser to Firefox recently... mainly for the very useful add ons available. One set of add ons I added has privacy in mind... one of these privacy focused add ons... was Multi-Acc Containers...}{Firefox} was labelled as a privacy post. 

\subsection{Training the Classifier to Identify if Users Discuss About Privacy in Software Product Forums}
\subsubsection{Model}
We trained a classifier to help us answer (RQ1) and to identify privacy related posts  that we could analyze for user privacy concerns (RQ2) and change in number of privacy discussion over time (RQ3).
Previous work using Reddit data \cite{iqbal_mining_2021} applied bag-of-words \cite{zhang2010understanding} and TF-IDF \cite{joachims1996probabilistic}. 
However, Devine et al. \cite{devine_evaluating_2021} compared text embedding techniques for analyzing user feedback and found that pre-trained deep learning models, particularly Universal Sentence Encoder (USE) \cite{cer2018universal}, performed much better than word frequency models like TF-IDF.
The model we trained using the base USE model with the transformer encoding mode \cite{use-url} had a precision, recall, and AUC of 0.84, 0.8, and 0.91. 
Privacy is not a common topic in app reviews or users posts, so our data was imbalanced.
Like previous work on Reddit \cite{iqbal_mining_2021}, we used the oversampling technique SMOTE \cite{smote2002} to address the imbalance.

\subsubsection{Manual Verification of Classified Results}
After running our USE model to identify whether users discuss privacy in software product forums via Reddit (RQ1), we randomly sampled 1000 posts from the data that contained 500 labelled privacy and 500 non-privacy. 
We manually labelled the 1000 posts to check the performance of our USE model.
We did not have access to the model's predictions until verifying the classified results was finished to reduce bias.
Our USE model achieved a precision, recall, and accuracy of 96.8\%, 96.8\%, and 96.7\% on the balanced random sample.

\subsection{Identifying the Privacy Concerns Users Discuss}
We clustered similar privacy posts to find main areas of privacy concerns to answer (RQ2).
We first grouped similar software together into categories as shown by Table \ref{clustering-categories}.
For (RQ2), we implemented clustering on the posts from each of the categories to find out the primary privacy concerns. Next, we found the best number of cluster using the silhouette coefficient as the metric before combining similar clusters across different categories into fewer clusters.
We used DistilBERT \cite{sanh2020distilbert} for embedding the data and performed dimension reduction using UMAP \cite{mcinnes_umap_2020} as part of clustering of our data. 
Moreover, UMAP is a general purpose machine learning dimension reduction algorithm that is both fast and scalable \cite{mcinnes_umap_2020}.
After dimension reduction, we conducted clustering via K-means \cite{k-means-info} and tried clustering with 2 to 10 clusters for each category. Clustering with more clusters (e.g. 20 clusters) can result in a larger number of privacy concerns, but our goal was to analyse the primary categories of privacy concerns so we choose 2 - 10 clusters.
We compared the results of different clustering and embedding using the silhouettes coefficient, which represents the distance of each point to the center of its cluster and with the closest neighboring cluster \cite{rousseeuw_silhouettes_1987}.
We randomly sampled a minimum of 25 posts from each cluster to compare with other clusters and merge similar clusters to collate common concerns. 

\begin{table}[h]
\caption{Categories used to group similar subreddits}
\begin{tabular}{@{}lp{6.5cm}@{}}
\toprule
\textbf{Category} &  \textbf{Subreddit} \\ \midrule
General Privacy & degoogle, privacy, PrivacyGuides, privacytoolsIO, \\
Social Media & Bumble, discordapp, facebook, facebookmessenger, Fiverr, instagram,   lineapp, linkedin, MicrosoftTeams, Pinterest, signal, SLACK, snapchat,   Telegram, Tinder, Twitter, Upwork, WeChat, whatsapp \\
Tools & androidapps, chrome, duckduckgo, duolingo, firefox, Google, kahoot,   miband, operabrowser, zoom \\
Platform & Android, chromeos, microsoft, ios, windows, windows8, windows10,   Windows11, \\
Entertainment & DisneyPlus, HBOMAX, netflix, soundcloud, spotify, youtube, YoutubeMusic \\
Financial & CoinBase, CashApp, venmo \\
Shopping & Aliexpress, amazon, Ebay, Wish \\
Food and Drink & deliveroos, doordash, McDonalds, starbucks, UberEats \\
Voice Assistant & amazonecho, googlehome, Siri \\
Travel & AirBnb, GoogleMaps, Lyft \\ \bottomrule
\end{tabular}
\label{clustering-categories}
\end{table}


\subsection{Mapping User Privacy Posts Over Time}
We mapped posts over time to observe the change in number of privacy discussion over time (RQ3).
We plotted the number of posts based on monthly time intervals.
Our rationale for plotting the number of posts over time was to observe if privacy regulations like the GDPR had any noticeable impacts on the number of privacy posts. 
Specifically, we wanted to visualize if there was an substantial increase of privacy posts when privacy regulations like the GDPR became law.
Finally, we normalized changes in privacy and non-privacy posts by their median as shown by Figure \ref{fig:normalized_total_posts}.

\section{Findings}
\subsection{RQ1: To what extent do users discuss privacy in software product forums? }
The results of applying our classifier are shown in Table \ref{results-privacy-count}: we identified 129,075 privacy related posts in our entire dataset of 4,488,467 total posts.
Privacy is represented by 2.9\% of all posts and 1.8\% if only considering software product subreddits.
1.8\% is higher than the 0.12\% found in prior work on Android app reviews \cite{nguyen_user_reviews_android}, but this was anticipated because users have more space to comment on multiple concerns in Reddit posts, and the Android sample was not focused on privacy.  
Popular social media software such as WhatsApp and Telegram not only had high numbers of privacy posts, but also had a high proportion of privacy posts.

\begin{table}[t!]
\caption{Result from classifying posts from all subreddits}
\centering
\begin{tabular}{@{}p{2.8cm}lrrl@{}}
\toprule
Subreddits & Classification                 & \multicolumn{1}{l}{Count} & \multicolumn{1}{l}{Percentage} &  \\ \midrule
62 Software Product Subreddits & \textbf{Non-Privacy} & 4274892                   & 98.2\%                         &  \\
& \textbf{Privacy}     & 78979                     & 1.8\%                          & \\ \midrule
4 General Privacy Subreddits & \textbf{Non-Privacy} & 134596                   & 96.3\%                         &  \\
& \textbf{Privacy}     & 50096                     & 3.7\%                          & \\ \midrule
& \textbf{Total Privacy} & 129075 & 2.9\% \\
& \textbf{Total Posts} & 4488467 & 100\% \\
\bottomrule

\end{tabular}
\label{results-privacy-count}
\end{table}

\subsection{RQ2: What privacy concerns do users discuss in software product forums?}
\label{results-privacy-concern}

\begin{table}[h]
\caption{Major privacy concerns and associated subreddit categories (from Table \ref{clustering-categories})}
\centering
\begin{tabular}{@{}p{2.5cm}|p{1.5cm}|p{3.6cm}@{}}
\toprule
\textbf{Privacy Concerns} &\textbf{Post Count}  & \textbf{Subreddit Categories}  \\ \midrule
Privacy Issues and Recommendations       &     39771 (31\%)    & (8) Entertainment, Tools, Food and Drink, Social Media,  Travel, Voice Assistant, Platform, General Privacy           \\
Privacy Policy and Permissions      &  27128 (21\%)     & (8) Tools, Financial, Shopping, Social Media, Travel, Voice Assistant, Platform, General Privacy               \\
Privacy Compromise Experiences     & 7037 (5\%)    & (5) Entertainment, Tools, Food and Drink, Financial, Platform         \\
Personal Information Exposures  & 23986 (19\%) &  (4) Food and Drink, Shopping, Social Media, General Privacy      \\
Unconfirmed Privacy Compromises    &     10715 (8\%) &    (4) Financial, Shopping, Social Media, Travel    \\
Bug and Suspicious Activity Complaints  &  15501 (12\%) &   (3) Entertainment, Financial, Shopping          \\
Warnings and Advisories    &  3829 (3\%)    & (3) Entertainment, Food and Drink, Travel   \\
Privacy Breach Effects   &   226 (0.2\%)     &  (2) Food and Drink, Travel            \\
Phishing Emails     &  885 (0.7\%)      & (1) Financial        \\
\bottomrule
\end{tabular}
\label{privacy-concern-clusters}
\end{table}

After embedding data using USE and then clustering the subreddit posts using K-means, we manually labeled those clusters and discovered that Reddit users primarily discuss 9 major privacy concerns. 

Table \ref{privacy-concern-clusters} shows the total number and name of the subreddit categories representing these privacy concerns.
The most common concerns among the categories were about \catname{software privacy policies and permissions}, and user experienced \catname{privacy issues and recommendations} for remedying the issues.
For instance, the post \qs{Apps which request sensitive permissions must provide link to valid privacy policy in the app and Google Play Developer Console}{Android} indicated the need of transparency for sensitive permissions. 

The next most common concern among subreddit categories was regarding users sharing their experiences and stories of \catname{privacy compromises} and scams describing the consequences of losing their privacy over the internet because of using a particular software product. 
To illustrate, \qs{I was hacked on my just eat account Saturday night someone ordered food and used my card details I had cleared my card details and sorting it out...}{Deliveroos}.

\catname{\small Unconfirmed privacy compromises} were instance where users discussed times they were unsure and suspected being hacked or scammed. 
A similar frequency was noticed for concerns where users are distressed about revealing their \catname{personal or monetary information} to the application and querying its safety on the forum. 
For instance, \qs{Quick question is it safe to put your debit/credit card info in? I have bought a few things from [AliExpress] before but used paypal... the seller i want to buy from does not accept paypal so before I buy I want to make sure its going to be safe.}{Aliexpress}.

Furthermore, complaints about privacy related \catname{bugs and suspicious account activities} were observed.
Users draws attention to some issues they were facing that were neglected by respective companies. 
For example, \qs{I've been dealing with this for several months now... I [struggle with Amazon], calling Amazon's customer service multiple times and asking if there is a way to force every logged in device to sign out of my account...Amazon should be able to secure my account... I wanted to... see if other people are dealing with it too.}{Amazon}. 
Next, we observed posts conveying \catname{warnings and advisories} for other users to avoid scams and even recommendations to avoid particular software to be secure from privacy breaches. 
For instance, the post \qs{Beware: WhtsApp web violation the privacy of Firefox Users. Use Telegram instead and uninstall this garbage!}{WhatsApp} specifically warned the other users of WhatsApp  terms and suggested to others to use other apps.

The remaining concerns were about the \catname{effects of privacy breaches} that represented the consequence of the privacy issue and violations of company rules and regulations by other users (such as illegal AirBnB cameras), and
user concerns about \catname{phishing emails} and email related scams. 


\subsection{RQ3: How do the number of privacy discussions change over time?}
\label{research_question_change}
\begin{figure}[t!]
    \centering
    \includegraphics[width=0.5\textwidth]{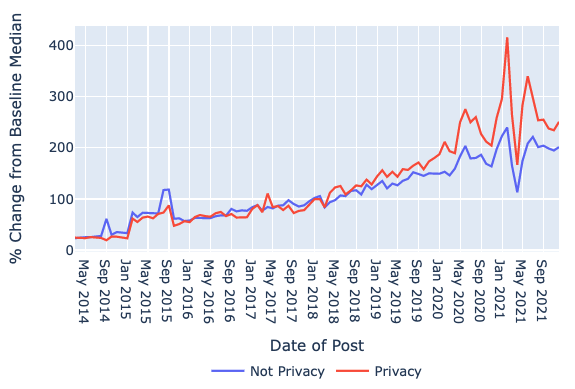}
    \caption{Change in all subreddits post count by month, normalized by the median post count for privacy and non-privacy.}
    \label{fig:normalized_total_posts}
\end{figure}

To analyze how the number privacy discussions changed over time
we mapped the posts over time and normalized the number of posts per month by the median post count starting from May 2014.
Although, we initially plotted the posts from the beginning of the subreddits' creation time, the number of posts before May 2014 were not significant enough to be shown on the graph. 
Thus, we showed the plot from May 2014 to 2021. 
We can further see from Figure \ref{fig:normalized_total_posts} that there was a steady increase in the number of privacy posts per month over time reaching a peak in January 2021.
In consideration of this peak, we observe a rise in the number of posts beginning in September 2020, so we calculate the growth rate for the 4 months leading up to January. 
Privacy posts showed a 119\% increase per month, compared to a 34\% increase per month for non-privacy posts, between September 2020 to January 2021.


\begin{figure*}
    \begin{subfigure}{.5\textwidth}

	    \centering
        \includegraphics[width=1\textwidth]{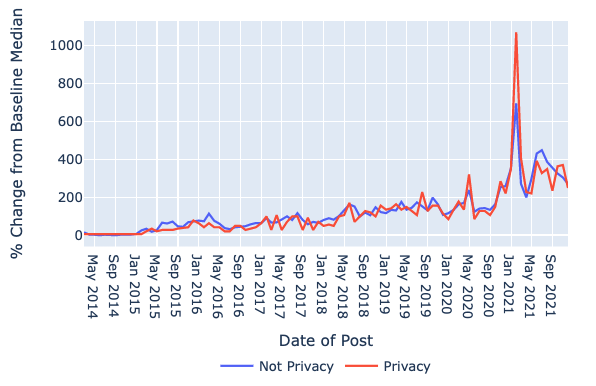}
        \caption{Telegram Subreddit}
        \label{fig:telegram-median-normalize}
	\end{subfigure}
    \begin{subfigure}{.5\textwidth}
        \centering
	    \includegraphics[width=1\textwidth]{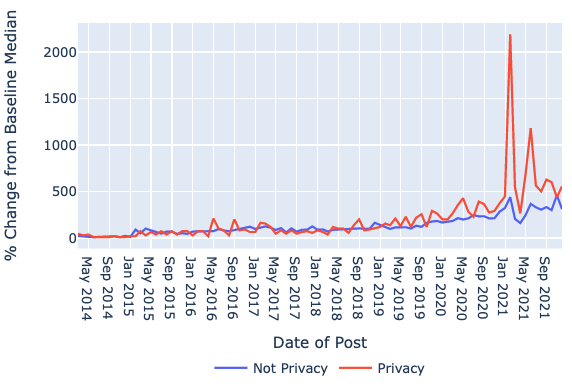}
        \caption{WhatsApp Subreddit}
        \label{fig:whatsapp-median-normalize}
   
	\end{subfigure}


\caption{Change in the number of posts per month, normalized by the median post count for privacy and non-privacy posts.}
\end{figure*}

Part of this increase can be explained by the overall increase in users posting on Reddit during this time, which aligns with reports that more users joined Reddit the last few years \cite{kastrenakes_Reddit_2020}.
Starting in January 2020 and culminating in January 2021, privacy related posts increased significantly over the median.
In January 2021, the number of privacy posts reached a new high of more than four times the median. 
While total posts on Reddit increased during this period, we can see from Figure \ref{fig:normalized_total_posts} that non-privacy posts did not increase in the same magnitude.

To better understand factors for the increases in the number of privacy discussions, we conducted a more thorough manual analysis of these posts. 
Our preliminary analysis indicates that an important factor for stimulating user privacy discourse are privacy \emph{narratives} that may trigger user behavior beyond the influence of privacy regulations. We describe this analysis next.

\section{Narratives as Explanations of Variations in Privacy Discussions}


\emph{Narratives} are a story or way of understanding a series of events that fosters a point of view \cite{hud4920}. 
We borrow the term {narratives} from economics \cite{shiller_narrative_2020} where researchers try to understand ``how stories go viral \& drive major economic events". 
They expand on the dictionary definition of ``narratives", a story involving humans, to include songs, jokes, and explanations.
In economic narratives research, people analyze historical events and have hypothesis on narratives that caused economic events.

For example, one of the sharpest economic US contractions ever occurred between 1920 to 1921 when consumer prices fell 16\%. 
One possible narrative explanation for the cause was that consumers were angry at supposed war profiteers in World War I and decided to boycott the profiteers by holding off buying necessary goods, thinking that they could get back at the profiteers \cite{shiller_narrative_2017}.
However, what people did not realize is that they would help cause a short term depression in the process.

For understanding the reason behind the outlier behaviour in the privacy discussion, we analyzed our data to see whether narratives can explain trends in the user discussions about privacy.
This is another area that a software organization can supplement as part of its requirements engineering processes.
It provides more understanding about the user's perspective on privacy. 
Moreover, for an organization this provides a whole new dimension for risk analysis before the development of a software product.  

We approached narrative analysis in two ways. One approach, top-down, looks for well-known events and identifies what narratives form in response. We chose the enactment of the GDPR as an example of this.
The other approach is to examine the data bottom-up and identify popular narratives that emerge from user discussions. We highlight narratives around chat apps Telegram and WhatsApp and browsing apps Chrome and Firefox as two examples.

\subsection{Introduction of the GDPR}
Although the number of privacy regulation posts was only a small fraction of our entire data, we found trends at various points in time with similar patterns for both events.

We found a total of 1351 posts that are related to the GDPR.
The first mention of GDPR was detected on 16 June 2015. 
However, the number of conversations about it was low before increasing in 2018.
Since the GDPR was enacted (came into effect) on 25 May 2018  \cite{radley-gardner_fundamental_2016}, we assume the GDPR enactment acted as a catalyst for the rise in posts during the deadline week (21-27 May 2018). 
78 posts mentioning GDPR were posted that week, out of which 22 were on 25 May, which is the maximum number of daily discussions on GDPR from 2015 to 2021. 
Observing Figure \ref{fig:normalized_total_posts}, we see that there is little change in the number of privacy posts in the  months around May 2018. 
The number of privacy posts is hovering around the median level, significantly lower than the level observed in 2020 or 2021.

These posts indicated user concerns over the use of their data by software like Google and Facebook. 
Posts like \qs{How can individuals get access to all their personal data being stored by companies like Google and Facebook... GDPR coming into force today entitles users access to their personal data... it would be fascinating to find out exactly what data is being stored and to what use it is being put...}{25 May 2018} highlight such concerns. 

However, we noticed a gradual drop in GDPR-related discussions as news surrounding the event wore down. 
Although the first GDPR related posts were identified in 2015, we found that the overwhelming majority of conversations regarding the GDPR occurred in the short window leading up to, and shortly after the privacy regulatory event. 
Moreover, most user discussions primarily focused on the privacy policies of big companies and user data concerns and not the regulation itself. 



Our analysis on posts after mapping them over time concerning privacy regulation indicates that \textbf{privacy regulatory events have a short term impact on people} (we report on the similar short-duration impact of the CCPA in our replication package).
Users mainly post about privacy regulations in the short interval during the week that the regulations became law. 

The enactment of the GDPR, while of short-term relevance, never acquired a viral status to persist beyond this period of time. We elaborate on the importance of the short-term narrative in the Discussion section.

\subsection{Telegram vs WhatsApp}
After Facebook announced plans to merge the infrastructure for WhatsApp, Instagram, Facebook, and Messenger \cite{perano_facebook_nodate}, the Reddit post data showed a outlier behaviour in posts from users in those related subreddits criticizing the move and expressing concern about the repercussions for their privacy and also comparing it with telegram.
We can see from Figure \ref{fig:whatsapp-median-normalize} and \ref{fig:telegram-median-normalize} that there were major outlier behaviour in the months near January 2021. 
In fact, the month of January 2021 represented a month of significant growth of the amount of privacy conversation from Whatsapp and Telegram. 
If software companies can identify the influence behind such increases, they can potentially link these concerns to privacy requirements and issues.

We thus manually analyzed 350 privacy posts from these five social media subreddits to get a deeper understanding. 
We found that there was a common theme between the privacy concerns for Telegram, WhatsApp, and to an extent other Facebook related software such as Instagram, Facebook, and Messenger and analysed these themes to understand the narratives behind.


\textbf{Narrative 1 - ``WhatsApp is sharing advertising data"}: 
Users voiced concerns regarding the implications from shared advertising data and merged infrastructure. For example, \qs{Now Facebook, Instagram and WhatsApp are all linked to pry on you...[they] are all now linked. How deep this could affect your privacy is not known}{Signal}.
The perception is that Facebook intends to integrate WhatsApp and other services, thereby increasing Facebook's advertising ability.
Users expressed concern for sharing advertising data, \qs{We talked about sweatshirts [sic] at WhatsApp with my friend now [I] am seeing sweatshirts [sic] ads everywhere on Instagram, [I] swear [I] did not googled or something else sweatshirts [sic] word}{Facebook}.

\textbf{Narrative 2 - ``Telegram is a privacy-centric alternative"}: 
We contrast the previous narrative with the narrative around Telegram, a WhatsApp competitor, in which the perception is that Telegram promotes user privacy.
The perception before and especially after Facebook's announcement is that a user could consider joining Telegram if they value their privacy, \qsd{[Signal] and Telegram are better than whatsapp. Yes my privacy is important to me...}{WhatsApp}{15 Jan 2021}.
Similarly, \qsd{Well, I've been using WhatsApp since 2010 and I honestly kinda wanted to quit for years but everyone I know kept using so what could I do? But with this new privacy policy, many of my friends have made the switch and I followed.}{WhatsApp}{11 Jan 2021}.
In addition to merging infrastructure, Facebook also introduced a mandatory and controversial privacy policy for WhatsApp \cite{damore_rachael_whatsapps_2021}.
A user must accept it to continue using WhatsApp \cite{damore_rachael_whatsapps_2021}. 
Narratives surrounding WhatsApp quickly became about WhatsApp's perceived privacy issues. 




However, not all users were convinced about Telegram's privacy.
For example, \qsd{After WhatsApp's new service policy, I see majority of my friends are moving to Telegram. They still believes It's more secure than Signal. I don't know why they do so. 
}{Signal}{23 Jan 2021} and
\qsd{The mark of Telegram is privacy, why only the secret chat has a good encryption?... if they boast so much about privacy, well they should show it, I don't feel so safe using the app anymore.}{Telegram}{20 June 2020}
Still, we see from Figure \ref{fig:telegram-median-normalize} that Telegram experienced about 1000\% increase in privacy and 700\% increase in non-privacy discussion from their median counts around the time of WhatsApp's controversial privacy policy.
Partially fueling this increase in user discussion was the massive increase in first time Telegram users.
Similarly, we see from Figure \ref{fig:whatsapp-median-normalize}, WhatsApp users discussing privacy a lot more than normal during this time. 
Narratives about these software seem like a potential driving force for the large spikes in user privacy discussion during this time, at a time when there was no other privacy regulation to influence privacy discussion.


In economics narratives, Shiller often compares narrative hypothesis with GDP growth and other economic measures to verify if narratives appear reasonable \cite{shiller_narrative_2017}.
We conducted a short comparison between the privacy policies between Telegram and WhatsApp to investigate whether the user concerns and privacy narratives regarding Telegram and WhatsApp are accurate representations of reality.

\textit{Privacy Policy Comparison Between Telegram and WhatsApp:}
We acknowledge that for a complete comparison we should compare the source code, infrastructure, and privacy policies. 
However, we found in Section \ref{results-privacy-concern} that discussion about privacy policy and software permissions is a primary user concern, indicating the importance that users place on a software's privacy policy.
One of our co-authors, a law scholar experienced in the intersection between law and software engineering, analyzed the respective privacy policies for WhatsApp \cite{whatsapp_privacy_policy} and Telegram \cite{telegram_privacy_policy}.   

In contrast to the privacy narrative fueling the drop in WhatsApp users and attracting millions of Telegram users \cite{dailey_telegram_2021}, their privacy policies demonstrates that user data is collected.
The privacy policies are complex legal documents filled with ``legal speak" that are not easy to understand for those unfamiliar with the jargon and those who do not have software engineering backgrounds. 
There are a host of technical terms that are foreign to lay persons.
At first, both privacy policies appear quite considerate of user data.
As per Section 3 of Telegram's privacy policy \cite{telegram_privacy_policy}, they do not collect a user's real name, but they collect phone number and email address as a backup. 
WhatsApp does not collect real names nor email, but it too collects phone number\cite{whatsapp_privacy_policy}. 
Both may collect location data, but they will do so only with user approval. 

Yet, later in the privacy policies, we saw statements that suggest that someone using the software may have their data collected. 
In Section 5.6 of Telegram's Privacy Policy  \cite{telegram_privacy_policy}, states that it does not collect one's data for ad targeting or other commercial purposes, but Section 6 of the same privacy policy states that 3rd party bots \emph{can} collect data. 
The statements seem to contradict each other and leaves the question open whether one's data is truly private on the platform.

WhatsApp's privacy policy also states that it may use and share information from other Facebook Companies and that information can be used for showing relevant offers and ads across the Facebook Company Products \cite{whatsapp_privacy_policy}.
WhatsApp's stated use and sharing of data seems consistent with the narrative {WhatsApp is sharing advertising data} and criticisms about the software's data sharing practices seem accurate. 
The perception around Telegram is that it will not share nor compromise user data with anyone, but according to Section 8 of its privacy policy, Telegram \emph{may} shares a user's personal data with companies in Telegram's parent group \cite{telegram_privacy_policy}.

This seems to contradict user perception, as many users are likely not aware that Telegram is collecting personal data at all, let alone sharing with other Telegram affiliated companies, which appears problematic for user privacy.
In contrast to the narratives and perception that users should switch to Telegram from WhatsApp, as Telegram is great for privacy and does not share advertising data, it is unknown whether a user can truly experience any benefits because \textbf{the two software systems have similar data sharing practices.}


\subsection{Firefox vs Chrome}
\textbf{Narrative 1 - ``Choose Firefox for its privacy":}
While Chrome is the more popular browser \cite{stat-counter}, we found that users frequently express concerns about Chrome's privacy and praise Firefox for its perceived privacy-centric approach for handling user privacy.
To this end, we found instances of users discussing their migration from Chrome to Firefox.
For example, 
\qs{A friend had told me about the privacy issues involved with Google Chrome so I decided to make the switch to Firefox.}{Firefox}
Our analysis indicates that users encouraged others to switch, which suggests that narratives could influence user perception and potentially even human behaviour to an extent. 

\textit{Privacy Policy Comparison Between Chrome and Firefox:}
We compared the respective privacy policies of Firefox and Chrome \cite{firefox_privacy_policy_2021, chrome_privacy_policy_2021} and the narratives that we observed from manual analysis.
Chrome's privacy policy stresses that ``You don't need to provide any personal information to use Chrome, but Chrome has different modes that you can use to change or improve your browsing experience. Privacy practices are different depending on the mode that you're using."
In other words, a user who uses Chrome only with the basic mode should in theory expect their browser data to be stored locally on their system \cite{chrome_privacy_policy_2021}.
However, if a user signs into Chrome with their Google Account, Chrome may offer to save a user's passwords, payment methods, and related information to their Google Account, though a user can turn off this setting.
Similarly, if a user uses features in Firefox beyond the defaults such as Firefox Suggest, what you type into the search bar and location information used to suggest content based on country, state, and city may be shared with Mozilla's (i.e. Firefox's parent organization) partners \cite{firefox_privacy_policy_2021}.

Privacy-centric browsing modes provide users with more options to address their privacy concerns, but we note that it is unclear the prevalence of users who truly understand the differences between the browsing modes and can decipher the use cases for when and what parts of their data may be shared. 
In 2019, Firefox began blocking third party trackers as a default setting as part of a major update \cite{oflaherty_firefox_2019}.
Although the setting was originally introduced in 2017, a mere 3\% of Firefox users applied the setting prior to the 2019 update, most likely because as suggested by a Firefox SVP, expecting users to alter their browser settings placed an ``undue burden" on them \cite{oflaherty_firefox_2019}.
Therefore, despite the narrative and perception that a user should switch to Firefox if they value their privacy, 
\textbf{it is unreasonable to assume that their privacy is secure without fully understanding the implications of each privacy setting.} 

Our preliminary analysis suggests that narratives surrounding software may play a profound role in shaping the perception and concerns about software products, and may even impact user behaviour. The privacy-centric narrative surrounding Telegram helped it increase its user count during the backlash against WhatsApp, at one point attracting 25 million new users in a span of 3 days  and reaching 500 million global users overall \cite{dailey_telegram_2021}. 
Similarly, the privacy-centric narrative regarding Firefox may have convinced some users to switch from Chrome to Firefox.

Narratives can be identified top-down---using external events to search for narratives---or bottom-up, browsing user discussions. 
Based on the preliminary study that we performed in this section, we conclude with this hypothesis for future studies:
\begin{mybox}[Hypothesis]
Privacy related narratives impact user privacy concerns about a software product. 
\end{mybox}

\section{Discussion}
Our study sought insights about privacy concerns in online communities, how privacy discussions vary over time and possible the causes for these  variations. 
We used Reddit as the source of data because the platform supports lengthy and detailed user discussions. 
We found that many Reddit users take advantage of the ability to engage in a community discussion about privacy concerns surrounding a wide variety of software (RQ1).
We found that the privacy issues noted by users can be broadly categorized into several groups including ``privacy compromise experiences" (RQ2), which often contain elaborate stories that users share about their difficulties encountered in a software that pertains to privacy. 
These privacy issues, some of which form bug reports and feature requests \cite{iqbal_mining_2021}, represent suggestions that may be developed into privacy requirements and inform organizations about significant issues at hand.

We identified that privacy regulations can influence the amount of privacy concerns over time, but the extent is limited to the short time period when laws are enacted. 
Our preliminary analysis on the contributing factors to variations in privacy discussion showed that privacy regulatory events are brief. 
We start our discussion by highlighting that users express many privacy concerns on Reddit and users often achieve this through elaborate and informative posts. 
Organizations can utilize user privacy concerns as these concerns represent needs and requirements from the users. An organization may consider and analyze these user privacy concerns as part of its requirements elicitation process. However, organizations should also reflect on the importance of privacy narratives, which we introduce in our preliminary analysis on causes of variations of privacy discussion, on the nature of the user discourse about privacy.

\subsection{Reddit: A Rich Source of Privacy Concerns}
Reddit provides a rich environment that supports online communities that thrive from open discussion and posts that encourage further discussion.
In contrast to other forums which have stringent length limits, such as Twitter, Google Play Store, and Apple App Store, Reddit allows user communities to discuss privacy concerns at length.
For some users, Reddit became the platform for them to share their personal stories about using a software and dissuade or persuade new users based on their experiences.

In one example, a user was able to explain in details about am incident where he was defrauded from a sizable amount of money via CashApp and was quite distraught from the whole ordeal. Not only was the user angry, but they told all their friends to stop using the app.
Users who experience this will often detail the ordeal on Reddit, which can generate negative publicity around the incident. 
It has been shown that users will refrain from using services when they do not feel that their privacy is safe \cite{balleste_pursuit_2013}, but we found that users may take further drastic measures if they feel their privacy is infringed.

We also observed users venting frustration about a new software feature that does not fit user requirements.
\qs{I recently transitioned to the `new' Facebook design as I assume most people did... I've always kept all elements of my Facebook profile viewable to only my friends or just myself... There is a ``public timeline" for my profile now with dozens of posts and photos and videos that are, apparently, visible to anyone on FB now... It's insane how complicated they've made this. I want to go back to having everything I post on FB be solely visible to my Friends...}{Facebook}

Part of improving product and service offerings for an organization involves identifying the users and collecting feedback from them to derive requirements \cite{groen_2015}.
Reddit is another avenue for collecting this feedback, much like app store reviews \cite{maalej_bug_2015, pagano_user_2013} or customer feedback. From the feedback, organizations could extract specific requirements based on criteria such as number of (angry) customers or fit with product roadmaps.
The addition of these relevant user privacy concerns to an organization's requirements elicitation process could assist the organization as it performs requirements analysis and develops privacy requirements from all the sources of feedback.


Our work brings empirical evidence that Reddit can be an important source for insights and information in the form of user discussion that organizations can leverage into developing into privacy related requirements.
However, we should note that we studied large, popular software products in this study and that we did not investigate smaller, less popular products, which may not have a wealth of user discussions. 
These smaller, less popular products may not attract the number of users, nor is the scale of user discussions about the products high, so our approach for analyzing Reddit posts may not translate for these smaller products. In addition to Reddit, we note there are other forums emerging daily (e.g., TikTok videos).

\subsection{Narratives: Unforeseen Influencer of Privacy Concerns}
To our surprise, objective privacy regulations for organizations had less impact on user perception about privacy than narratives.
We found that privacy related narratives impacted user perceptions about software systems and may have contributed to an increase in users moving from one software to another.
The impact of narratives on human behavior is not new \cite{wydick_bruce_how_2015}.
We found in our work that many users believed in the narrative that Telegram is more privacy conscious than WhatsApp (and found something similar for Firefox vs Chrome).

Although our exploratory analysis of the respective privacy policies indicates that narratives may not be accurate in representing reality, at least not to the extent that many users are led to believe, we see correlations with privacy concerns and these narratives.
The spread of narratives also shares similarities with memes, where knowledge is spread and copied to other individuals \cite{heylighen_f_memetics_2001}, but predicting the influence of narratives is hard \cite{shiller_narrative_2017}.

Narratives can spread quickly \cite{shiller_narrative_2017}, 
as in the cases where Whatsapp was going to share advertising data and Telegram was great for privacy. 
Users that believed such narratives and think that WhatsApp collects user data for profit may decide to find alternative software.
We cannot state that WhatsApp or Telegram is more private than the other without a full comparison of every aspect about the software, but our qualitative analysis on their privacy policies shows that their data collection practices are similar.
However, Telegram gained millions of users in part due to the ``Whatsapp is sharing advertising data" and ``Telegram is a privacy-centric alternative" narratives \cite{dailey_telegram_2021}. 

These narratives can have far reaching consequences that heavily impact an organization's business as Meta (i.e., WhatsApp and Facebook's renamed parent company) suffered its first drop in daily users in its 18 year history during 2021 \cite{dibenedetto_meta_2022}.
The organization's stock also dropped more than 20\% in a single day, which was one of the single largest drops to a single company in history \cite{dwoskin_facebook_2022}
There are likely multiple factors that contributed to this large drop in valuation and daily users that extend beyond solely privacy concerns.
Focusing exclusively on traditional requirements gathering and product management approaches is not sufficient: the impact of user concerns and perceptions also strongly influence the popularity of a software and company reputations.
If an organization is able to identify a negative privacy narrative from user discussion, the organization's best course of action may be to consider strategies or factors to mitigate the privacy narrative. 
One strategy may involve developing privacy requirements that aim to address the privacy narrative. 
Alternatively, an organization may choose to employ branding or advertising to diminish the effects of a negative privacy narrative. 
Exploring how organizations manage privacy narratives is considered future work and we elaborate more in the research implications section.


As observed by our analysis on user concerns about the GDPR, some narratives such as privacy regulatory events, may have a short term impact on users. 
A small number of users may hear about the privacy regulation in the media and large organizations go offline as a result of the regulations  \cite{lecher_major_2018}, causing user concerns for a brief period of time when the regulation becomes law.
Then user interest and relevance about the privacy regulatory narrative quickly wanes, despite the large implications that such regulations could have on software organizations. 

Ultimately, privacy narratives have non-negligible impact on user perceptions around software. 
Our work not only shows the potential for organizations to leverage user feedback into privacy requirements, but also underscores the important role that narratives play in shaping privacy concerns. 
Since privacy narratives can impact privacy concerns in a way that is subjective from reality, organizations could benefit from considering and analyzing privacy narratives so that they could maximize the insights elicited from user feedback.
Considering and understanding privacy narratives could be a windfall for an organization as it elicits privacy requirements from users.

\subsection{Research Implications}
We believe there are several research implications for future work.
First, although we identified a series of user privacy concerns from user discussions on Reddit, developing privacy requirements from these user privacy concerns was not part of our study. Future work can explore deriving privacy requirements, perhaps through automated measures, based on user discussions in online user communities. 
Moreover, our work found plenty of privacy discussion from users of large, popular software products, though future work could investigate if the same is true for smaller, less popular software.
In particular, what is the threshold for the amount of user feedback a less popular software product needs before we can develop narratives for the software.
Furthermore, in the future we can utilize different features of the posts in our dataset to better understand the relevance of privacy
discussions and the change of a topic over time.
Likewise, further research can investigate how narratives form for less popular software products and how those narratives impact the products. 

Based on our preliminary analysis we hypothesize that ``privacy related narratives impact user privacy concerns about a software product".
This hypothesis should be tested in future work involving a deeper analysis on the impacts of privacy narratives. 
Next, further research is necessary to develop metrics or strategies to systematically identify narratives in user feedback in the requirements engineering context.
More importantly, additional work is needed to understand identify the narratives that become viral and significantly impact an organization's software systems.
Finally, further investigation on \emph{if} and \emph{how} organizations consider and manage \emph{narratives} are important areas for future work. 
In particular, we need to further study how we can incorporate identification and analysis of narratives as part of an organization's requirements elicitation process to more effectively achieve an understanding of user requirements.


\subsection{Practitioner Implications}
Our work indicates that organizations could take advantage of analyzing user discussions on Reddit as it is a platform where rich discussions about software can take place. 
An organization can elicit and learn about privacy concerns from analyzing user discussions in the organization's corresponding subreddits.
These privacy concerns may be further refined and developed into privacy requirements that an organization can realize in the software.
In addition, organizations should be cognizant of narratives that could shape user perceptions and concerns about privacy.
As demonstrated in our work, user behaviour and their concerns could be influenced by narratives, which may not always be an accurate reflection of reality.
Hence, an organization should allocate resources in considering these macro level narratives that could greatly impact an organization's software. 
Since a narrative represents an aggregate of user feedback, an organization can develop requirements from a narrative similar to how it may develop requirements from several user feedback at a time. 
Our work illustrates that an organization may expand its requirements elicitation process to include the addition of user discussion data from Reddit and analyzing it for privacy narratives. 
The added insights from the requirements elicitation may facilitate improvements to the organization's understanding of user privacy concerns and help it develop more relevant privacy requirements. 
Finally, further investigation with respect to the actual users' understanding is needed to understand the scope and validity of different narratives.

\section{Threats to Validity}
\subsubsection{External validity}
The first threat is that generalizability of our results could be limited because we collected only 66 subreddits. 
However, to mitigate this we collected the subreddit data from a wide range of applications and services. 
Our data is comprised with subreddits that are associated with popular software products such as AirBnb, or Instagram which have relatively higher user counts, but our data also contain popular software products with fewer number of user posts. 
Since our focus was on popular software products that generally have large numbers of users, it is possible that our approach for identifying user privacy concerns and narratives does not generalize well to software products that do not have a high number of user discussions. 

\subsubsection{Construct validity} The threat applies to the manual labeling of posts to privacy and non-privacy which is to prepare our ground-truth data.
Manual labeling can cause experimental bias as humans tend to be subjective in their judgment which can be very difficult to eliminate, but text classification is generally done manually. 
We tried to address this problem by having two experts who are well versed, experienced, and understand privacy. 
We calculated our Cohen’s kappa value and the inter-rater agreement levels which also reflects our labeling efficiency. 
A similar issue of subjectivity also applies when it comes to analyzing and comparing privacy policies as lawyers have different judgments of privacy policy implications. 
For example, lawyers had conflicts in the interpretation of complaints in GDPR regulations \cite{li_towards_2022}. 
The possibility of different interpretations cannot be eliminated. 
We tried to mitigate this threat by having a law scholar compare these policies. 

\subsubsection{Internal Validity} 
There are threats internal validity that relate to our understanding of the data.
It was not possible for us to manually cluster each post to privacy concerns.
To mitigate this threat we took our 9 main areas of privacy concerns and randomly sampled at least 25 posts from each area to check if they belong. 
Furthermore, there are limitations to our choice of time intervals for mapping user privacy posts over time.
A researcher would need to reduce the time interval to daily or hourly if they want to visualize short-term narratives. 
Contrastingly, one may need to expand the time interval to yearly if they want to visualize the impact of long-term narratives. 
We also acknowledge that finding all impacts of privacy regulations on user concerns is not possible.
To address this we extracted all the combinations of terms for GDPR, collected the relevant posts, and manually identified if a post is talking about these regulations. 
One threat to our work is that we did not take advantage of addition attributes corresponding to a post such as number of comments or number of up votes. As our primary focus was answering our RQs we did not investigate the effect from the additional information, we leave that for future work.
However, we address that we could not explore other aspects of the privacy regulatory events such as considering the impact of changes made in the products due to the enactment of the policies that could potentially trigger the user discussion.
Another potential limitation of our work is our interpretations of narratives.
Our study of narratives was exploratory and our definition originated from economics. 
We acknowledge that our approach of identifying narratives from observing multiple similar user posts and matching these with news stories from the same time frame may not be the best approach, but we believe our work still brings attention to this area of research.
Future work may leverage tools or strategies from social network analysis \cite{golbeck} for further study of narratives.

\section{Related Work}
\subsection{Privacy Related Requirements Elicitation}
The enactment of privacy laws like GDPR and CCPA influence organizations to link these laws to privacy requirements for their software \cite{breaux_privacy_2014}. 
Breaux states that organizations can derive value from improving privacy requirements and creating a more personable experience for users \cite{breaux_privacy_2014}. Islam et al. developed the foundation of a framework to identify and map relevant legal terms to privacy requirements  \cite{islam_towards_2010}. However, these privacy laws are notoriously complex and have ambiguous language \cite{breaux_analyzing_2008}. 
Bhatia et al. empirically analyzed these privacy policies using Tregex Patterns
\cite{bhatia_automated_2016}, and Reidenberg et al. proposed a natural language processing (NLP) technique to improve the clarity of the policies so companies can use them \cite{reidenberg_ambiguity_2016}. 
Nonetheless, privacy requirements elicited from these laws are not enough to address user concerns \cite{gharib_privacy_2016}. 
Some case studies have focused on eliciting user privacy requirements \cite{kalloniatis_addressing_2008, gharib_privacy_2016}. 

There have been several studies that have examined user perceptions regarding privacy and privacy policies \cite{sheth_us_2014, norberg_privacy_2007, brown_studying_2001}.
Unlike studies that investigated developing privacy requirements or requirements elicitation, these studies focused on the human aspects of privacy.
Of particular relevance to our work is the notion of the ``privacy paradox" \cite{brown_studying_2001}, in which users complain about privacy (as in our paper) yet willingly trade their privacy for minor advantages.
Whether our narratives reflect this paradox we leave for future work, although our observance of WhatsApp's decreasing and Telegram's increasing user numbers indicates that some connection may exist between user concern and behaviour.  


\subsection{Eliciting Requirements from App Reviews}
Numerous studies focused on improving the requirement engineering process by understanding users’ perspectives from the reviews on  app marketplaces and user forums \cite{tizard_can_2019, pagano_user_2013, kanchev_social_2015, williams_mining_2017}. 
Pagano and Maleej \cite{pagano_user_2013} identified the patterns, topics, and quality of user feedback in over one million Apple App Store reviews and studied its impact on the software requirements. 
Prior work has shown the importance of user feedback from software forums \cite{tizard_can_2019} and social media platforms \cite{kanchev_social_2015, williams_mining_2017, iqbal_mining_2021}. 
Tizard et al. \cite{tizard_can_2019} analyzed user reviews and feedback from the two product forums (VLC and Firefox) and concluded that product forums are a valuable source of consumer feedback that is essential for the evolution of the product. 
For social media, Kanchev et al. \cite{kanchev_social_2015} performed a preliminary analysis of  user discussions on Google maps from Reddit discussions and presented examples of requirement-related artifacts. 

A recent study by Iqbal et al. \cite{iqbal_mining_2021} analyzed characteristics of Reddit posts about software applications and found out that more than half of the posts contains useful information such as bug reports and feature requests.

Automating feature extraction from app stores to find crowd-based requirements engineering was looked at in \cite{guzman_how_2014, harman_app_2012}.
For example, Guzman and Maalej \cite{guzman_how_2014} proposed to identify fine-grained app features by using collocations and sentiment analysis about the identified features and grouping them using topic modeling into a more meaningful high-level feature. 

Recent research include different approaches for the automatic processing of user reviews \cite{chen_ar-miner_2014, carreno_analysis_2013, maalej_bug_2015}. 
Since a significant amount of work has been done on App Reviews, we focused on Reddit to analyze users' privacy concerns and identified influence of factors like narratives on the discussions. 

\subsection{Narratives}
Nobel Prize winning professor Robert J. Shiller explains 
that one's mind can be swayed by “narratives”, even if these stories or trends are unfactual \cite{shiller_narrative_2017}. 
An example was when various governments struggled to persuade citizens to adopt COVID-19 tracing apps \cite{ng_contact-tracing_nodate}, despite the theoretical benefits. 
Bano et al. analyzed app reviews and news articles and found that though governments tried to impose the COVID-19 contact-tracing apps for the benefit of their citizens, governments did not anticipate the socio-political-cultural factors that caused failure of the apps \cite{bano_rise_2021}.
It supports Shiller's point that we need more research into narratives and how it impact human behavior \cite{shiller_narrative_2017}. 
Nevertheless, stories (i.e., narratives) with purpose can play powerful roles in understanding customer needs, allowing product innovation \cite{snowden_story_1999}.

To the best of our knowledge, we are the first to study privacy concerns among users on forums and their temporal evolution to observe fluctuations over time using Reddit as a source.
We found that there were inexplicable fluctuations, leading us to analyze narratives as an important factor for privacy concerns. 

\section{Conclusion}
Analyzing user discussions on social media platforms has become an essential aspect of requirement elicitation. 
We focused on analyzing user discussions on software product forums to investigate users privacy concerns and the number of their concerns over time.
We used Reddit, a host of many online communities for software products, as the source of our data. 
Our approach to classifying privacy related posts and clustering them into nine main areas helped us identify the types of privacy related conversations. 
We identified that privacy concerns are influenced by narratives.
Our study shows that an organization can gather insights about user privacy concerns from user discussions on Reddit to strengthen the organization's requirements elicitation; an organization may also develop privacy requirements from the insights about user privacy concerns as requirements analysis and development is the next step following requirements elicitation. 
Our work was exploratory so it may be worthwhile for future work to comprehensively explore other developer communities.
Researchers can examine the discrepancy between user and developer posts in other software forums regarding privacy concerns. 
Furthermore, studies can explore methods to consider the extent of influence of narratives on privacy concerns and other research implications related to narratives.

\bibliographystyle{IEEEtran}
\bibliography{main}

\end{document}